\documentclass[journal,hideappendix]{vgtc}        % final (journal style) without appendices
\vgtccategory{Research}

\title{Context-KG: Context-Aware Knowledge Graph Visualization with User
Preferences and Ontological Guidance}

\author{%
  Rumali Perera$^{1}$, 
  Xiaoqi Wang$^{2}$, 
  Han-wei Shen$^{1}$ \\[12pt]
  $^{1}$The Ohio State University \qquad $^{2}$Bosch AI Research
}
\authorfooter{
  \item
  	Rumali Perera is with The Ohio State University.
  	E-mail: perera.62@osu.edu.
  \item
  	Xiaoqi Wang is with Bosch Center for Artificial Intelligence.
  	E-mail: wang.5502@osu.edu.

  \item Han-Wei Shen is with The Ohio State University.
  	E-mail: shen.94@osu.edu.
}

\abstract{%
Knowledge Graphs (KGs) are increasingly used to represent and explore complex, interconnected data across diverse domains. However, existing KG visualization systems remain limited because they fail to provide the context of user questions. They typically return only the direct query results and arrange them with force‑directed layouts by treating the graph as purely topological. Such approaches overlook user preferences, ignore ontological distances and semantics, and provide no explanation for node placement. To address these challenges, we propose Context-KG, a context-aware KG visualization framework. Context-KG reframes KG visualization around ontology, context, and user intent. Using Large Language Models (LLMs), it iteratively extracts user preferences from natural language questions and context descriptions, identifying relevant node types, attributes, and contextual relations. These preferences drive a semantically interpretable, ontology-guided layout that is tailored to each query, producing type-aware regions. Context-KG also generates high-level insights unavailable in traditional methods, opening new avenues for effective KG exploration. Evaluations on real world KGs and a comprehensive user study demonstrate improved interpretability, relevance, and task performance, establishing Context-KG as a new paradigm for KG visualization.
}

\keywords{Human-centered computing, information visualization, visual analytics, graph drawing.}

\teaser{
 \includegraphics[width=0.9\linewidth]{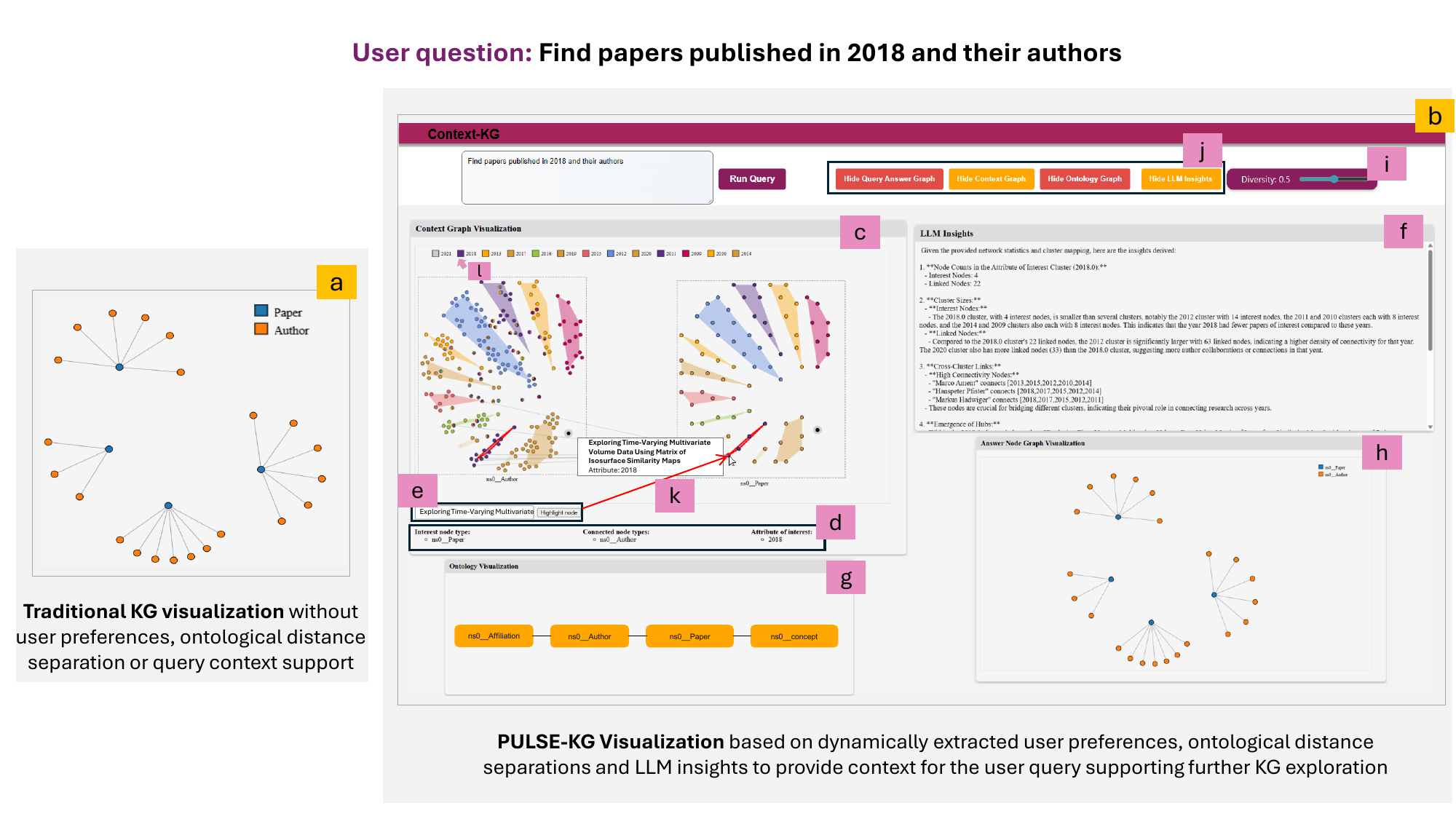}
 \centering
  \caption{User query run on the TVCG KG using (a) a traditional force-directed layout and (b) Context-KG, which incorporates dynamically extracted user preferences, ontological distances, and insights for context-aware exploration.}
\label{fig:Teaser}
}

\graphicspath{{figs/}{figures/}{pictures/}{images/}{./}} % where to search for the images

\usepackage{tabu}                      % only used for the table example
\usepackage{booktabs}                  % only used for the table example
\usepackage{lipsum}                    % used to generate placeholder text
\usepackage{mwe}                       % used to generate placeholder figures
\usepackage{ccicons}                   % package to be able to use icons from creative commons

\usepackage{mathptmx}                  % use matching math font
\usepackage{makecell}
\usepackage{multirow}
\usepackage{array} 
\usepackage{amsmath}
\usepackage{soul}
\usepackage{algorithm}
\usepackage{algorithmicx}
\usepackage{algpseudocode}

\usepackage{xcolor}

\usepackage{amsmath,amsfonts,bm}

\def\eqref#1{equation~\ref{#1}}
\def\1{\bm{1}}

\DeclareMathAlphabet{\mathsfit}{\encodingdefault}{\sfdefault}{m}{sl}
\SetMathAlphabet{\mathsfit}{bold}{\encodingdefault}{\sfdefault}{bx}{n}

\usepackage{amssymb}
\usepackage{amsmath,amsfonts}
\usepackage{amsopn}
\usepackage{bm} % bold symbol
\usepackage{multirow}
\usepackage{flushend}
\usepackage{tabularx}
\usepackage{xspace}
\def\eqref#1{equation~\ref{#1}}
\def\1{\bm{1}}
\DeclareMathAlphabet{\mathsfit}{\encodingdefault}{\sfdefault}{m}{sl}
\SetMathAlphabet{\mathsfit}{bold}{\encodingdefault}{\sfdefault}{bx}{n}
\newcommand{\eat}[1]{}

\providecommand{\manuscriptnotetxt}{}
\renewcommand{\manuscriptnotetxt}{}
\begin{document}

%%%%%%%%%%%%%%%%%%%%%%%%%%%%%%%%%%%%%%%%%%%%%%%%%%%%%%%%%%%%%%%%
%%%%%%%%%%%%%%%%%%%%%% START OF THE PAPER %%%%%%%%%%%%%%%%%%%%%%
%%%%%%%%%%%%%%%%%%%%%%%%%%%%%%%%%%%%%%%%%%%%%%%%%%%%%%%%%%%%%%%%

% \firstsection{Introduction}

\maketitle

\section{Introduction}
\label{intro_sec}

Knowledge Graphs (KGs) have become the foundational infrastructure for modern information organization, offering expressive graph structures and ontological semantics that support complex reasoning. However, KGs remain notoriously difficult for humans to explore and interact with effectively. The power of KGs is locked behind query languages such as SPARQL or Cypher that inherently assume deep familiarity with the underlying schema, ontology, and data distribution. Users who lack this expertise often fail to extract meaningful insights, not because the KG lacks the answers, but because formulating the “right” query requires knowledge they do not yet possess. This leads to a persistent paradox: users must understand the KG to query it, but must query the KG to understand it. This limitation is amplified in KG visualizations. Most existing systems render raw triples with flat, semantically opaque layouts that provide little contextual guidance, making it hard to identify meaningful regions, relationships, or promising avenues for exploration. Thus, users misinterpret proximity, overlook important connections, and struggle to build a coherent mental model of the KG.

\begin{figure}[h]% specify a combination of t, b, p, or h for top, bottom, on its own page, or here
  \centering % avoid the use of \begin{center}...\end{center} and use \centering instead (more compact)
  \includegraphics[width=\columnwidth, alt={Comparison of visualizations for the query “Given four papers, what are some of their concepts and who are the authors?” in (a) Neo4j which places concept and author nodes near unrelated papers, creating misleading proximity and (b) our ontology-aware layout separates node types into distinct regions, preventing overlap between unrelated entities.}]{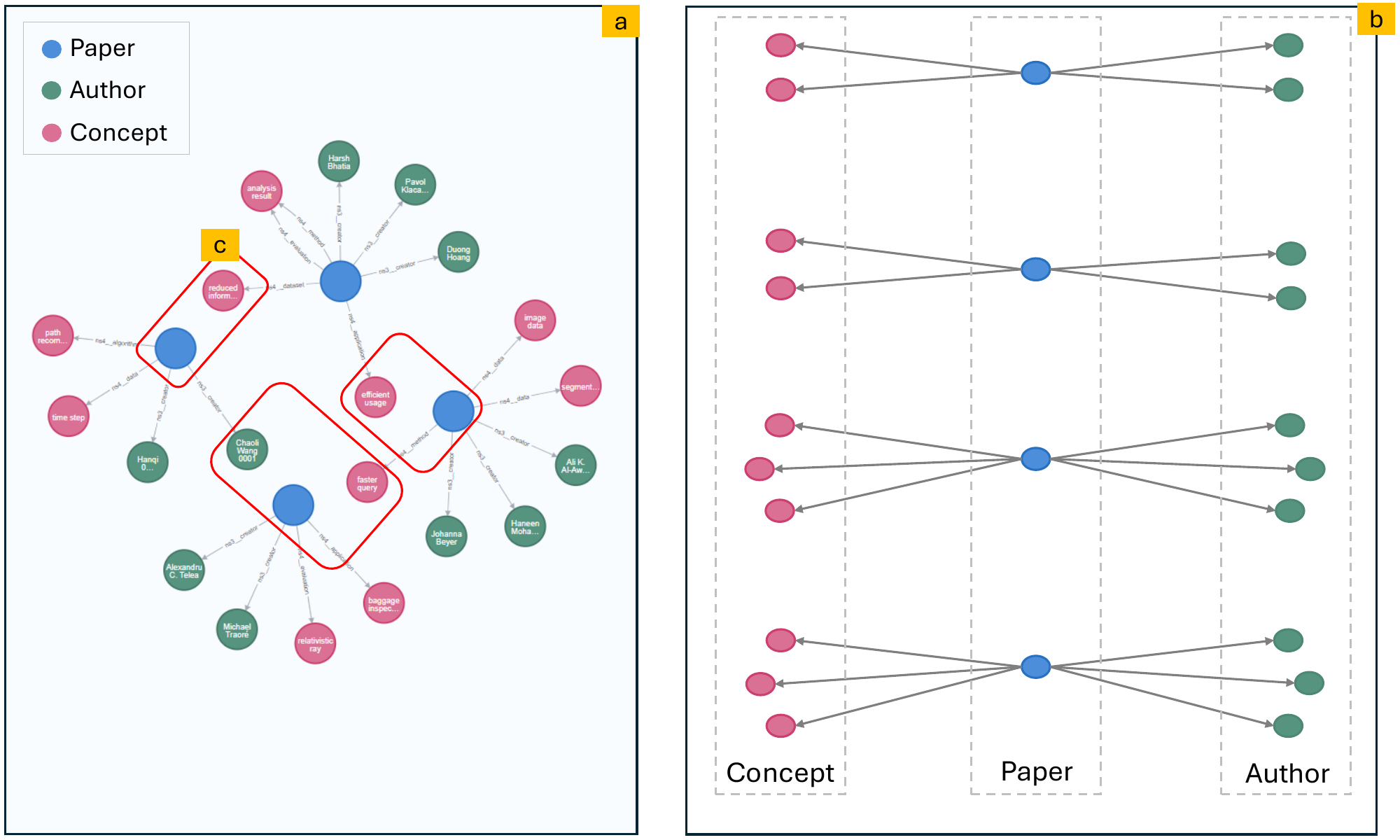}
  \caption{%
  	Comparison of visualizations for the query “Given four papers, what are some of their concepts and who are the authors?” in (a) Neo4j which places concept and author nodes near unrelated papers, creating misleading proximity and (b) our ontology-aware layout separates node types into distinct regions, preventing overlap between unrelated entities.%
  }
  \label{fig:node_types}
\end{figure}

A central factor behind this problem is that KGs encode semantic and ontological structure, but widely used tools such as Neo4j \cite{kemper2015beginning} and KingVisher \cite{both2024kingvisher} rely on force-directed layouts (e.g., Fruchterman–Reingold \cite{fruchterman1991graph}, Kamada–Kawai \cite{kamada1989algorithm}). They treat nodes only as particles under pairwise forces and ignore these semantic layers and ontological structures. This causes unrelated entities to appear close together and semantically related ones to scatter, producing misleading spatial cues that distort user understanding \cite{saket2014node, pfeffer2013fundamentals, mcgrath2004you}. For example, consider the query drawn from the TVCG KG \cite{tu2024kg}:
“Given four papers, what are some of their concepts and who are the authors?”
Conventional force-directed layouts (Figure~\ref{fig:node_types}, Region a) scatter concepts, papers, and authors, placing unrelated nodes adjacent to one another simply to satisfy force equilibrium. By contrast, an ontology-aware layout (Figure~\ref{fig:node_types}, Region b) clearly separates semantic regions, preserves type-based neighborhoods, and prevents visual misconceptions. This is not merely an aesthetic improvement, it is fundamental to preventing cognitive bias and semantic misinterpretation. 

Furthermore, existing KG visualizations also neglect the fact that users hold unique, dynamic preferences \cite{qiu2024unveiling}. A domain expert, a student, and a data engineer querying the same KG may desire qualitatively different contextual information and yet no existing system adapts to these differences. Current preference-aware KG visualization methods require either a manually selected seed node \cite{lissandrini2022knowledge}, which oversimplifies real-world information needs, or extensive historical behavioral logs \cite{liu2024knowledge}, which are seldom available, difficult to gather, and not suited to cold-start scenarios. Moreover, none of these approaches extract user preference signals directly from natural language questions, a critical gap highlighted in recent surveys \cite{cheng2025survey, li2023knowledge}. This makes current KG systems unable to provide per-query personalization, leaving all users with identical, one-size-fits-all layouts. Because preference extraction is typically cast as a fixed-label classification problem \cite{bouza2014partial}, these methods fail on vague, evolving, or multi-stage intents \cite{si2020learning} and cannot capture open-ended, context-dependent information-seeking. Thus, the core problem is three fold: 1) current KG visualizations fail to provide contextual guidance for exploration, 2) they do not incorporate user preferences and 3) they lack emphasis on the ontological structure of the graph.

Motivated by these critical gaps, we propose Context-KG, a user-preference-driven, context-aware KG visualization framework that integrates three core components: (1) Large Language Model (LLM) driven iterative user preference extraction, (2) ontology-aware spatial layout, and (3) automated insight generation. Given a user’s natural language query and additional context statements, Context-KG interprets user intent using an LLM, maps it onto the KG’s ontology, and produces a layout that clusters semantically related nodes. Interest nodes (main node types mentioned in the user’s query) are grouped based on attributes of interest, and other relevant nodes are positioned relative to these clusters according to the ontological structure. Users can iteratively refine context using the additional context statements, which the system translates into visual modifications (Neighbor, Edge, or Path Context). The resulting visualization goes beyond triple rendering by generating an ontology-grounded semantic map with clearer conceptual regions, interpretable spatial structure, and explainable node placement. Finally, the LLM generates interpretive insights summarizing patterns, anomalies and contextual relevance in the Context-KG aiding users in knowledge discovery.
Context-KG is a novel framework to unify iterative LLM-driven preference reasoning, ontology-grounded spatial layout, and automatic insight generation into a single KG visualization paradigm.
% To the best of our knowledge, Context-KG is \shen{tone down: the first framework} to unify iterative LLM-driven preference reasoning, ontology-grounded spatial layout, and automatic insight generation into a single KG visualization paradigm. 
Each component is not a minor variation of existing techniques; rather a novelty in how the LLM, ontology, and iterative context visualization are jointly integrated into a context-aware and user-adaptive KG exploration workflow. The main contributions of our work are as follows:

\begin{itemize}

\item A novel LLM-driven method that interprets natural language questions and context descriptions to iteratively infer user preferences for KG visualization.

\item An ontology-aware context KG visualization that leverages class hierarchies and semantic structure to produce cognitively aligned spatial layouts, mitigating misleading node placements.

\item An insight-generation module that derives explanations, summaries, and exploration guidance from the context visualization, bridging graph rendering and actionable understanding.

\item A unified visual analytics system that operationalizes the full pipeline, combining preference extraction, ontology-aware context layout, interactive exploration, and automated insight generation to substantially enhance KG usability.
\end{itemize}

\section{Related Work}
\label{sec:related_work}

Our work is related to the following three fields: KG visualization, Focus$+$Context in graph visualization and force-directed graph layouts which are summarized in this section.

\subsection{Knowledge graph visualization}
Early KG visualization explored self-organizing maps on spherical surfaces \cite{zhu2013knowledge} as an alternative to Euclidean layouts. Systems such as Neo4j and KinGVisher provide graph-based views of nodes and relationships. VizKG \cite{raissya2021vizkg} offers a framework for visualizing SPARQL query results over KGs, featuring a wide range of visualization types and recommendation capabilities. KGViz \cite{nararatwong2020knowledge} utilizes local open-source LLMs to construct KGs from unstructured text by offering interactive visualizations and RDF triple exports. Thinkbase and Thinkpedia \cite{hirsch2009interactive} aim to make web content more accessible by creating interactive graph-based representations of large knowledge repositories like Freebase \cite{bollacker2008freebase} and Wikipedia.

However, these approaches lack question-specific contextualization and offer only static, non-adaptive views. They provide limited support for interpreting results within broader attribute space or adjusting to evolving user intent. In contrast, our approach dynamically reconfigures visualizations based on extracted user preferences and ontological context, ensuring that users not only see the immediate query results, but also gain contextualized views.

\subsection{Focus+Context in Graph Visualization}

Focus$+$Context techniques in graph visualization provide detailed views of selected areas (focus) while preserving awareness of the overall structure (context), thereby supporting exploration and reducing cognitive overload. Early approaches, such as fisheye views and bifocal lenses, enabled smooth navigation between detail and overview \cite{furnas1986generalized}. Later work introduced volumetric focus$+$Context methods for interactive navigation in domains such as medical imaging and geospatial analysis \cite{zhao2013volumetric}. The concept of \textit{augmented context} highlighted multiple objects of interest within the contextual view using spatial strategies and visual cues to minimize disruptive pop-ups \cite{huot2007focus+}. More recent approaches use multi-focus probes that reveal subgraphs in detail while maintaining clear visual connections to the overall network \cite{zimmermann2025multi}. 

However, existing focus$+$Context techniques for KG visualization seldom integrate user preferences or attribute-driven clustering, reducing their ability to adapt to evolving exploration goals or highlight user-relevant patterns. They also typically ignore ontological structure, leading to misleading spatial organization. Our work addresses these limitations by coupling user preference extraction with attribute-based clustering and ontology-aware layouts to produce more interpretable, question-aligned context views.

\subsection{Force Directed Graph Layouts}

The foundation for modern force-directed graph layout approaches was laid by Eades, who introduced the spring layout method \cite{eades1984heuristic}. This model was later refined by Fruchterman and Reingold, improving force distribution to enhance spatial uniformity and clarity \cite{fruchterman1991graph}. Another notable advancement is the Kamada-Kawai algorithm, which calculates forces based on graph-theoretic distances to ensure that node placement reflects the shortest paths in the graph \cite{kamada1989algorithm}. More recently, the t-FDP model \cite{zhong2023force} has been introduced, incorporating a bounded short-range force based on Student’s t-distribution, leading to improved neighborhood preservation and enhanced performance for large-scale graphs. These methods also account for graph aesthetics to enhance readability and clarity \cite{wang2021deepgd,wang2023smartgd}.

However, all of these existing methods primarily focus on the structural aspects of graph layouts ignoring node semantics and ontological distances, limiting their effectiveness for KG visualization. To address this, our work introduces a semantically and ontological distance aware force-directed context KG layout, enabling more interpretable and semantically meaningful visualizations.

\section{Design Requirements}
\label{sec:design_requirements}

Based on previous studies and surveys on visualization and KG exploration \cite{li2023knowledge}, \cite{riva2024visualization}, \cite{qiu2024unveiling}, \cite{chen2022exploring}, \cite{jeong2025geokg}, \cite{law2020data}  we identify a set of design requirements for Context-KG.

\textit{R1. Providing context for user questions.} 
KG systems must supply contextual cues linking user questions to relevant underlying data, ensuring that non-experts can interpret retrieved entities and their relationships \cite{li2023knowledge}.

\textit{R2. User preference extraction.}  \label{req:R2}
Capturing user preferences from natural language to align with the graph’s structure and semantics \cite{chen2022exploring}, especially when users lack formal query expertise.

\textit{R3. Ontological Structure Preservation.} \label{req:R3}
Nodes should be arranged by ontological hierarchies to preserve semantic relationships and clarify both direct and indirect connections \cite{tu2024kg}.

\textit{R4. Explainability of Node Placement for Visual Inference.} 
Clearly convey why nodes are positioned as they are: based on semantic similarity, ontological distance, or connectivity. This enables users to understand relationships, trace connections, and draw accurate inferences.

\textit{R5. Interactive Exploration.} \label{req:R5}
Support iterative refinement of user questions and provide responsive visual updates that facilitate deeper exploration \cite{li2023knowledge}.

\section{Methodology}
\label{sec:method}

\begin{figure*}% specify a combination of t, b, p, or h for top, bottom, on its own page, or here
  \centering % avoid the use of \begin{center}...\end{center} and use \centering instead (more compact)
  \includegraphics[width=\textwidth, height=7cm, keepaspectratio, alt={The overview of Context-KG. User question is processed by an LLM for preference extraction. The extracted preferences are processed by the LLM with KG schema information to extract the KG instances. The resulting nodes of interest are clustered by the user-selected attribute and arranged along an arc, with the diversity parameter controlling the visualization scope. Connected nodes are positioned according to ontological distances, while the LLM further supports insight generation.}]{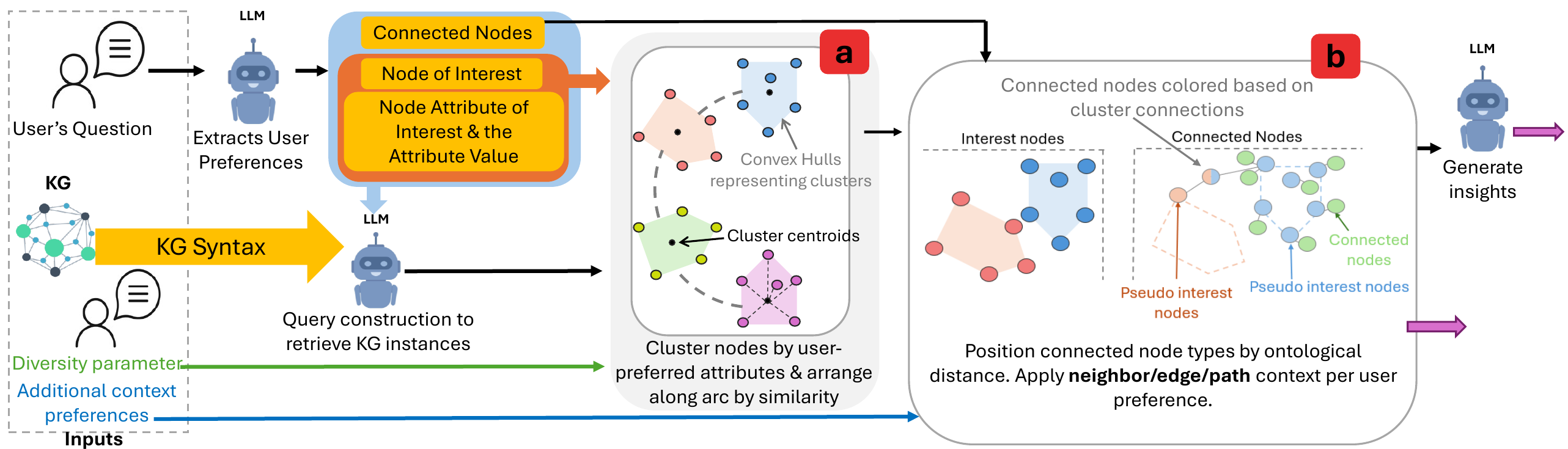}
  \caption{%
  	% \shen{the text in the figure are way too small} 
    The Context-KG overview: the user question is processed by an LLM to extract preferences, which are combined with KG schema to identify relevant instances. The resulting nodes of interest are clustered by the user-preferred attribute and arranged along an arc, with the diversity parameter controlling the visualization scope. Connected nodes are positioned according to ontological distances, while the LLM further supports insight generation.
    % \shen{I found this figure **very** hard to understand} 
    %
  }
  \label{fig:overview4}
\end{figure*}

This section introduces our Context-KG framework, designed to generate context-aware KG visualizations tailored to user questions. Our method follows a three-stage pipeline: (1) understanding user preferences from natural-language questions and context descriptions, (2) producing a context-driven visual layout, and (3) generating interpretable insights via LLMs. Figure~\ref{fig:overview4} provides an overview of the system. We use the following terminology for clarity: a "node" is an individual entity (e.g., a paper, author, or organization in the TVCG KG), a "node type" is the class of that entity (e.g., Paper, Author, Affiliation in the TVCG KG), and the "ontology" represents the KG schema, where node types are vertices and edges represent schema-level relationships.

\subsection{User Preference Understanding}
\label{subsec:user-preferences}
Our framework begins by interpreting the user’s natural-language question to uncover their intent goals and establish an initial context for visualization. Since understanding intent from free-form language requires semantic reasoning beyond keyword or rule-based techniques, we rely on an LLM to interpret user questions \cite{han2024beyond}. LLMs can generalize across paraphrases, resolve ambiguity, infer implicit preferences, and align everyday expressions with the KG schema \cite{bodonhelyi2024user, liu2024monotonic, tsaknakis2025llms}. Building on this interpretation, we systematically represent each user intent through four key elements extracted from their natural language question: (1) the primary node type of interest (2) the key attribute of that node type, (3) the attribute value specified in the question, and (4) other mentioned node types, which we refer to as \textit{connected node types}. Table \ref{tab:user_preferences} provides example user questions that can be asked from different types of KGs and the resulting four key elements extracted as user preferences. 
% \shen{Is it the eurovis style that you do not need to say Fig. in 1? It reads wierd to me though}
By concentrating on a single primary node type and its associated attribute, we reduce cognitive load and support targeted contextual reasoning \cite{li2023knowledge, spritzer2011design}. Using these extracted preferences (\hyperref[req:R2]{R2}), the LLM generates schema-aware queries \cite{liang2021querying}, \cite{liu2025survey} which are executed on the KG to retrieve the relevant nodes and relationships for the visualizations.

The user’s primary natural language question forms only the starting point of the visualization. Users can iteratively provide additional context descriptions as input to enhance the context visualization (\hyperref[req:R5]{R5}). The LLM classifies these inputs into one of three context types, Neighbor Context, Edge Context, or Path Context, drawing on the GAKE framework \cite{feng2016gake}. Neighbor Context emphasizes important nodes, Edge Context identifies critical relationships, and Path Context highlights informative connections between nodes. For Path Contexts, users can select from three criteria shortest paths \cite{madkour2017survey}, homogeneous paths \cite{guttmann2020relational}, or disjoint paths \cite{kzeinberg1995disjoint}, to surface meaningful structural patterns in the KG \cite{madkour2017survey, guttmann2020relational, kzeinberg1995disjoint}.

\begin{table}[t]
\centering
\footnotesize
\caption{Examples of context descriptions across different KG types and their corresponding context types}
\label{tab:context_examples}
\setlength{\tabcolsep}{6pt}
\renewcommand{\arraystretch}{1.1}
% \resizebox{\columnwidth}{!}{
\begin{tabular}{|p{1cm}|p{3.9cm}|p{2cm}|}
\hline
\textbf{KG Type} & \textbf{Context Description} & \textbf{Context Type} \\  
\hline

Academic KG
& Show me which of these authors are the most prolific. 
& Neighbor Context \\
\cline{2-3}
& Highlight edges representing first-author contributions. 
& Edge Context \\
\cline{2-3}
& Show how Paper A is connected to Affiliation B? 
& Path Context — Shortest Path \\
\hline

Smart Foodshed KG
& Which of these food hubs are the most connected in the whole supply chain? 
& Neighbor Context \\
\cline{2-3}
& Which Organizations are the lead organizations of the Projects? 
& Edge Context \\
\cline{2-3}
& Show purchase paths where all products belong to the same category. 
& Path Context — Path Homogeneity \\
\hline

% \multirow{3}{*}{\textit{Social Media KG}} 
% & Which users are the most influential in this community? 
% & Neighbor Context \\
% \cline{2-3}
% & Highlight edges that represent direct message interactions. 
% & Edge Context \\
% \cline{2-3}
% & Identify user-to-user paths that minimize shared intermediary accounts. 
% & Path Context — Path Disjointness \\
% \hline

\end{tabular}
% }
\end{table}

\begin{table*}[t]
\centering
\caption{User preferences extracted from user questions using LLM}
\label{tab:user_preferences}
\setlength{\tabcolsep}{6pt} % Adjust the column separation
\renewcommand{\arraystretch}{1.1} % Adjust the row height
\resizebox{\textwidth}{!}{
\begin{tabular}{|c|c|c|c|c|c|}
\hline
\textbf{KG Type} & \textbf{User Question} & \textbf{Node type of interest} & \textbf{Attribute of interest} & \textbf{The value of the attribute of interests} & \textbf{Connected node types} \\  
\hline
\textit{Academic KG} & Show Tier A conferences in computer science, their papers and authors & Conference & Tier & Tier A & Paper, Author \\
\hline
\textit{Social Media KG} & Find users who joined in 2021, their friends and the groups they belong to & User & Joined year & 2021 & Friend, Group \\
\hline
\textit{E-Commerce KG} & Which orders were placed for handbags from Brand X, and who were their customers? & Order & Product Brand & Brand X & Customer \\
\hline
\textit{Movie KG} & Show actors above age 40, their released movies and the movie directors & Actor & Age & 40 & Movie, Director \\
\hline

\end{tabular}
}
\end{table*}

\subsection{User-Centric, Context-Aware KG Visualization}

\subsubsection{Ontology-Guided Global Arrangement}
\label{subsec:ontology-layout}

% \shen{you can probably start by definnig the terminolgy - a node, a node type, an ontology node, etc. The word nodes are used all over the place but they mean differently in different context. It is very difficult to switch the understadning of node meannig as you do bedause they are trying to understand your method so they not have what you have in mind in their mind. So from the user point of view, you are placnig nodes, cluster nodes, link connected nodes etc. This is a disaster.When in the introduction you do not describe your main idea and only say what you can do, and in the method section you cannot describe your idea clearly, this will make it impossible for you to receive a good rating}

% The visualization establishes a global spatial layout where node types are positioned according to the KG’s ontological structure.
% % \shen{what is a node} 
% Treating the ontology as a graph with node types as vertices and relations as edges, the visualization space is partitioned into regions corresponding to these types. 

To provide a meaningful overview of the KG, the visualization first establishes a global layout that reflects the ontology structure (\hyperref[req:R3]{R3}). Node types are arranged spatially based on their relationships in the ontology, creating intuitive regions that capture semantic organization. This ontological arrangement ensures that closely related node types appear near each other, while more distant types occupy separate regions, supporting a high-level understanding of the KG’s overall structure. By modeling the ontology as a graph with node types as vertices and relations as edges, the visualization leverages pairwise shortest-path distances between types to guide placement. Formally, let $D = [d_{xy}]$ denote the pairwise shortest path distances between node types $x$ and $y$ in the ontology. We define this distance as the \textit{ontological distance}. Node type positions $p_x$ and $p_y$ are then arranged to satisfy:

\begin{align}
\| p_x - p_y \| \approx d_{xy}
\end{align}

ensuring that Euclidean distances in the visualization approximate the ontological separation between node types. Since most real-world ontologies can be formulated as connected graphs, this structure generalizes well.

\begin{algorithm}[htbp]
\footnotesize
\caption{User-Centric, Context-Aware KG Visualization}
\label{alg:context-kg-visualization}
\begin{algorithmic}[1]
\Statex \textbf{Input:} $KG=(N,E)$ with $N$ nodes and $E$ edges, Ontology $O=(T,R)$ with $T$ node types and $R$ relations, User preferences $P_{\text{user}}$, Context descriptions $\mathcal{C}$

\Statex \textbf{Output:} Visualization $V$, Insights $I$

\State \textbf{1: Ontology-Guided Layout:} Build ontology graph $O_G=(T,R)$, Compute distance matrix $D = [d_{xy}]$, $d_{xy}=\text{dist}(t_x,t_y)$, Assign positions $p$, s.t. $\|p(t_x)-p(t_y)\|\approx d_{xy}$
\State Partition layout space into disjoint regions $\{R_t \mid t \in T\}$

\State \textbf{Semantically-Aware Node Organization:}
\State Extract interest node $I \subseteq N$ and attribute $A$ from $P_{\text{user}}$
\If{$A \in \mathbb{R}$}
    \State $C \gets \text{DBSCAN}(X)$; label cluster $c \in C$ by $\mu(A_c)$
\ElsIf{$A \in \text{Text}$}
    \State $E \gets \text{Embed}(A(I))$, $C \gets \text{K-means}(E,k)$; label $C$ via $\text{LLM}$, $k$ choosen via WCSS
\EndIf

\If{node $n \in I$}
    \State Sample $I_s \subseteq I$ with diversity $\sigma$
    \State Place cluster centroids on circular arc, position $n$ radially; anchor with pseudo-edges applying repulsion/attraction, enclose cluster with convex hull
\Else\; $N_c \gets $ Connected nodes
    \State Initialize $x \in N_c$ near pseudo-interest node anchors $P^\ast$
    \State Position $x$ at centroid if linked to multiple interest nodes
    \State Color $x$ for multi-cluster links; visualize as pie chart
\EndIf

\State \textbf{Step 3: Incorporate Additional Context Descriptions}
\State Classify $\mathcal{C}$ into {Neighbor, Edge, Path} via LLM
\If{$c$ is Neighbor Context}
\State Compute scores for nodes and adjust node sizes
\ElsIf{$c$ is Edge Context}
\State Highlight edges matching relationship attributes
\ElsIf{$c$ is Path Context}
\State Select path criterion (shortest, homogeneous, or disjoint) and highlight relevant paths between specified nodes
\EndIf

\State \textbf{Step 4: Intelligent Insights Generation}
\State Encode features $\Phi(V)=\{ \text{deg}(n),\text{clusters},\text{bridges},\text{outliers}\}$
\State $I \gets \text{Query LLM}(P_{\text{user}}, O, \Phi(V))$
\State Return $(V,I)$
\end{algorithmic}
\end{algorithm}

\subsubsection{Semantic-Aware Node Organization}
\label{subsec:node-clustering}
In this section we describe how the layout is created within the ontology-defined regions and how it is refined iteratively with context descriptions.

\begin{figure}[h]% specify a combination of t, b, p, or h for top, bottom, on its own page, or here
  \centering % avoid the use of \begin{center}...\end{center} and use \centering instead (more compact)
  \includegraphics[width=\columnwidth, alt={Node layout showing (a) connected nodes positioned via pseudo interest nodes, (b) clusters with convex hulls, and (c) the layout without convex hulls depicting clusters.}]{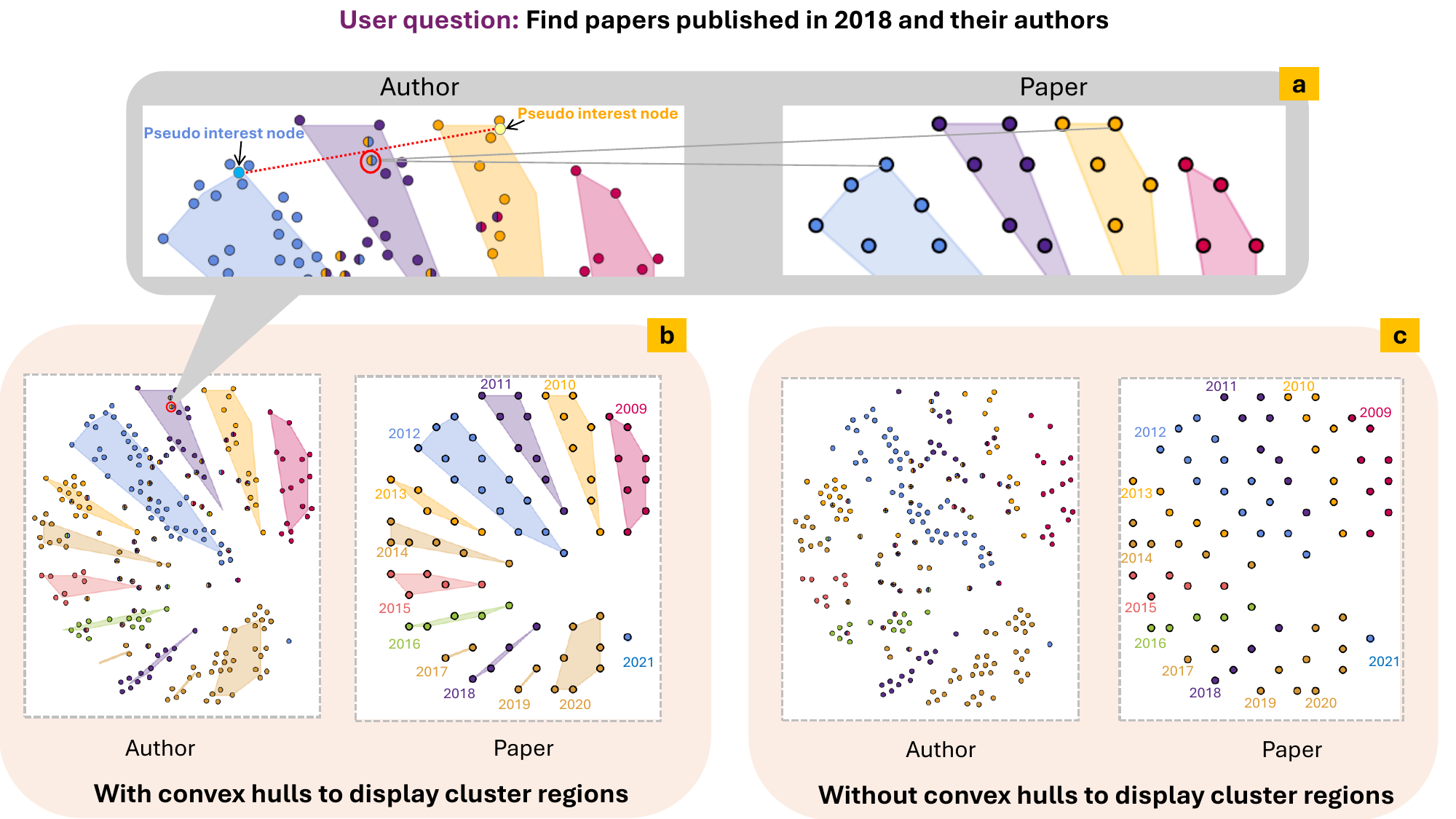}
  \caption{%
  	Node layout showing (a) connected nodes positioned via pseudo interest nodes, (b) clusters with convex hulls, and (c) the layout without convex hulls depicting clusters.
  }
  \label{fig:Convex_Hull_Position}
\end{figure}

\textbf{Spatial Organization of Interest Nodes.}
% \shen{I found the term 'node' confusing. }
To help users interpret complex KGs, the visualization organizes nodes so that meaningful patterns become immediately apparent. Within the ontology-defined regions, the visualization groups nodes of interest by clustering, to reveal meaningful patterns. This also helps mitigate the unstructured “hairball” effect common in KG visualizations. By clustering nodes of interest according to user-preferred attributes, the system surfaces temporal, thematic, or structural patterns. This allows users to move beyond individual nodes and observe higher-level structures in the data. For instance, in Figure~\ref{fig:Convex_Hull_Position}, Region b, the query “Find papers published in 2018 and their authors.” sets papers as the node type of interest and year as the attribute, producing year-based clusters shown as colored polygons in the \textit{paper} region. For Numeric attribute clustering we used the DBSCAN algorithm\cite{ester1996density}, which handles noise and non-convex shapes well. Textual attributes were clustered using the K-means algorithm. DBSCAN is well-suited for numeric data but unstable for high-dimensional text embeddings, while K-means handles text embeddings with roughly convex clusters but is sensitive to numeric outliers. Clusters were labeled using mean values for numeric attributes or LLM-generated cluster topics for textual attributes. Real-world KGs contain millions of nodes, making it infeasible to display all clustered %\shen{again, what is the "nodes" here?} 
nodes. To address this, we sample nodes using a user-adjustable diversity parameter \cite{nachmanson2015graphmaps}. Sampling begins from the cluster containing the user-preferred attribute value. High diversity samples proportionally from all clusters and low diversity emphasizes the cluster with the preferred attribute and its neighbors. Prior work has shown that incorporating diversity improves insight discovery \cite{lv2025boosting}. 

After clustering the nodes of interest, the visualization positions them to convey both cluster structure and inter-node relationships in a clear and interpretable manner. Cluster centroids were first arranged along a circular arc based on semantic similarity (Figure~\ref{fig:overview4}, Region a), ensuring that related clusters are close while avoiding misleading proximity between distant clusters.% \shen{I found this confusing - first of all, again, is eurovis's style not to use 'Fig' or 'Figure' in 3-a? Seocond, typically 3-a means a sub figure but you do not have figure 3-a. 3-a is only a region. Very confusing} 
We avoided straight-line placement which would  cause occlusions when placing connected nodes and fully circular arrangements were avoided, as node clusters at opposite ends could appear adjacent, misleading interpretation of distant clusters. Interest nodes within each cluster were positioned radially, proportional to their number of connected nodes to reduce overlap. Formally, the radial distance $r_i$ of node $i$ from the cluster centroid was computed as: 

\begin{align}
r_i &= r_{\min} + \frac{c_i}{c_{\max}} \, (r_{\max} - r_{\min})
\end{align}

where $c_i$ is the maximum linked node count for node $i$, $c_{\max}$ is the maximum linked node count among all nodes within the cluster, and $r_{\text{min}}$ and $r_{\text{max}}$ define the minimum and maximum allowed radii for positioning nodes within the cluster region.

\begin{figure}[h]% specify a combination of t, b, p, or h for top, bottom, on its own page, or here
  \centering % avoid the use of \begin{center}...\end{center} and use \centering instead (more compact)
  \includegraphics[width=\columnwidth, alt={How the use of the diversity parameter changes nodes of interest sampling from clusters based on a node threshold}]{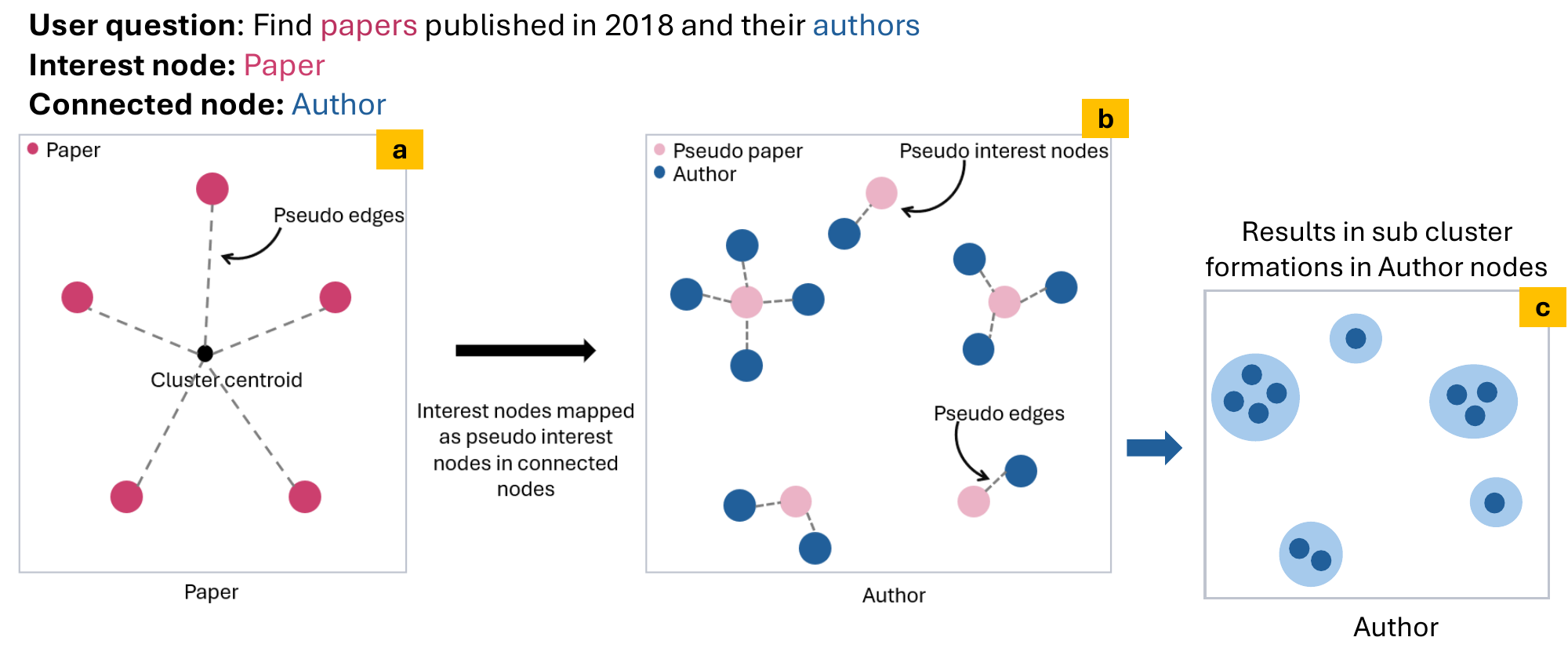}
  \caption{%
  	(a) Psuedo-edges for anchoring each node of interest to its cluster centroid, (b) Pseudo-interest nodes linked to the connected nodes and (c) as a result of (b), sub clusters forming in connected node regions  %
  }
  \label{fig:PseudoNodesAndEdges}
\end{figure}

To maintain cluster cohesion, we introduce a novel refinement of the classical Fruchterman–Reingold algorithm using pseudo-edges (Figure~\ref{fig:PseudoNodesAndEdges}, Region a) that anchor each node of interest to its cluster centroid. In this refinement, repulsive forces act between all nodes to prevent overlaps, while attractive forces are mediated through these pseudo-edges. This centroid-anchoring mechanism explicitly encodes cluster structure to ensure that local cohesion is preserved without sacrificing global readability. To visually distinguish clusters, distinct colors were assigned and their regions was marked using convex hulls. Without this the visualization would reduce to scattered points, offering little spatial context. This can be seen in Figure~\ref{fig:Convex_Hull_Position}, Region c (no convex hulls to represent clusters) compared to Figure~\ref{fig:Convex_Hull_Position}, Region b (clusters outlined with convex hulls).

% \begin{figure}[h]% specify a combination of t, b, p, or h for top, bottom, on its own page, or here
%   \centering % avoid the use of \begin{center}...\end{center} and use \centering instead (more compact)
%   \includegraphics[width=\columnwidth, alt={Shortest path distances between node types defined as the ontological distance and their respective placement in the context visualization}]{fig/Ontological_Distance}
%   \caption{%
%   \shen{text in figure way too small, and what are stream/publication/creator?}
%   	Shortest path distances between node types are defined as the ontological distance and their respective placement in the context visualization. This is a sketch of the DBLP KG \cite{ackermann2024dblp} ontology showing the core entities.% 
%   }
%   \label{fig:Ontological_Distance}
% \end{figure}

\textbf{Spatial Organization of Connected Nodes.} 
To make the relationships between interest nodes and their connected nodes visually interpretable, we organize connected nodes relative to the clustering of interest nodes. This ensures that connected nodes naturally form sub-clusters that reflect multi-hop patterns, making it easier to observe how groups of interest nodes relate to their neighborhoods across the KG. For this purpose, we organize connected nodes  using the modified Fruchterman–Reingold layout, anchored by pseudo-interest nodes (Figure~\ref{fig:PseudoNodesAndEdges}, Region b), which mirror each real interest node and serve as fixed attractors. Connected nodes were initialized near these anchors and iteratively updated using global repulsion and edge-based attraction to their pseudo-interest nodes, producing a spatially coherent and uncluttered distribution. This resulted in the formation of sub-clusters within the connected node type regions (Figure~\ref{fig:PseudoNodesAndEdges}, Region c), as connected nodes linked to a single interest node naturally group together. When a connected node is linked to multiple interest nodes across different clusters, it is positioned near the centroid of the forces exerted by these nodes. This reflects its multi-cluster associations Figure~\ref{fig:Convex_Hull_Position}, Region a. To enhance node placement explainability, connected nodes were displayed as pie charts, with each wedge representing the proportion of links to different interest node clusters. This makes multi-cluster connectivity immediately interpretable \cite{saket2018task}. 

% \shen{my overall comment about your method section is it is very hard to read. The section significantly lacks clarity. Your figures are not effective, and the explaination lacks high to low level description. It is probably a passnig grade but it has a lot to desire. I have hard time to fully understand the term 'node'. For example, you say you are clustering node of interests but you say node is an ontology node, say paper. If the meaning of a node is 'paper', an ontology type, why do you need to cluster them. There is only one paper node type in ontology so I am not sure what you are clustering. If you mean you are cluserting papers, then the meaning of node is not a node paper in the ontology. This really hinder the reading }

\begin{figure}[h]% specify a combination of t, b, p, or h for top, bottom, on its own page, or here
  \centering % avoid the use of \begin{center}...\end{center} and use \centering instead (more compact)
  \includegraphics[width=\columnwidth, alt={Iterative context refinement in Context-KG: (a) edge context, (b) path context, (c) neighbor context. }]{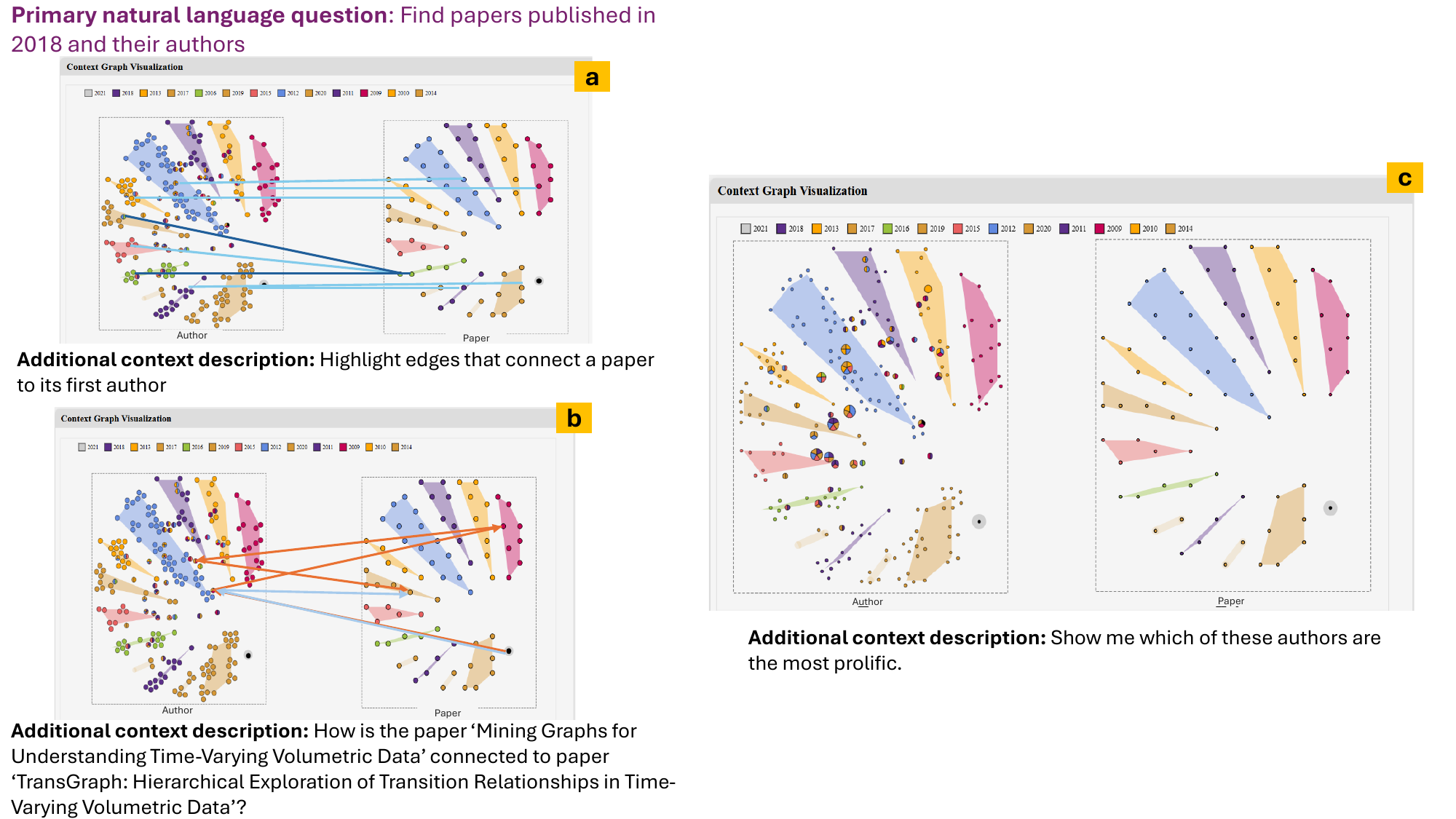}
  \caption{%
  	Iterative context refinement in Context-KG: (a) edge context, (b) path context, (c) neighbor context.%
  }
  \label{fig:context_additional}
\end{figure}

\textbf{Iterative Context Descriptions for Layout Refinement.} To make the visualization more responsive to user intent, we allow iterative updates that reflect additional context provided during exploration. Crucially, these context layers are made possible by the use of an LLM, which interprets under specified or ambiguous user requests and translates them into precise visualization modifications, capabilities beyond what rule-based methods can achieve.
Based on the classification of context descriptions as neighbor context, edge context or path context, the LLM is used to identify what visual changes need to be applied to get the context updates given the ontology summary, node of interest, attribute of interest and connected nodes. Neighbor Context emphasizes important nodes via size (Figure ~\ref{fig:context_additional}, Region c). In the Edge Context (Figure ~\ref{fig:context_additional}, Region a), we determine which edge attributes warrant emphasis. To reduce visual clutter, we use edge bundling, allowing users to selectively expand bundles on demand, revealing only the edges of interest. Path Context identifies all paths between source and target nodes (Figure ~\ref{fig:context_additional}, Region b).
The LLM prompts used in this section are provided in the supplementary material. The complete procedure of the Context-KG visualization is summarized in Algorithm ~\ref{alg:context-kg-visualization}.

\textbf{Context-KG Layout Parameters.}

Context-KG layout algorithm utilizes fixed iteration counts and temperature annealing as implicit convergence mechanisms to ensure layout stability. The main layout runs for 100 iterations \cite{davidson1996drawing} with an initial temperature $T_0$ set to one eighth of the squared diameter of the layout region. The temperature decays multiplicatively \cite{fruchterman1991graph} by 0.94 each iteration to gradually reduce node displacement. Connected nodes are anchored over 80 iterations with a reduced temperature of one-fifth $T_0$. Node overlap resolution runs up to 10 iterations, stopping early if nodes stabilize.

\subsubsection{Intelligent Insights Generation}
\label{subsec:llm-insights}
%Intelligent Insights Summary
To support users of varying expertise \cite{li2023knowledge}, we enhance Context-KG with LLM-generated insights. We first encode the structural and relational features of the context KG visualization, including how nodes are organized into clusters and how they connect across different node types. This encoding is incorporated into a carefully designed prompt that guides the LLM to generate comparative insights tailored to the user's question. The full version of this prompt is provided in the supplementary material. In doing so, the system translates the raw context data into natural language explanations that highlight insights such as hubs, cross cluster connections, cluster sizes, degree distributions, bridging nodes and outlier clusters. The full set of LLM insights for the query “Find papers published in 2018 and their authors” in Figure~\ref{fig:Teaser}, Region f, can be found in the supplementary materials.

% \subsubsection{Design Logic of the Visualization}
% Our design choices stem from three integrated considerations. First, conventional KG visualizations risk misleading users by placing all nodes in a single space where visual proximity implies false semantic relatedness. To avoid this, we structure nodes into ontologically defined regions, to ensure that spatial proximity reflects their conceptual distances. Second, our design emphasizes the node type of interest by clustering and arranging them along a circular arc, with convex hulls depicting the cluster boundaries. This clustering-centric view provides a way for users to interpret how other connected node types are positioned relative to these primary clusters, supporting exploratory “deep dives” aligned with user-driven investigative questions. Third, our approach aids explainability in node placement (R4). Interest nodes occupy positions influenced by their clusters, while connected nodes are placed according to their connections with the interest nodes but in their ontological distance aware regions. We further apply a color-coding scheme, reflecting the clusters to which they link. This encoding reveals higher-order structural signals, such as nodes that bridge multiple clusters, nodes limited to a single cluster, and the density distribution of connected memberships within clusters. These explainability-oriented design choices align with growing needs in the KG visualization literature.

\subsection{Visual Analytics System}
\label{subsec:va-system}

We develop a Visual Analytics system Figure~\ref{fig:Teaser}, Region b,  to demonstrate the effectiveness and usability of our proposed pipeline. The interface was built on an infinite canvas with modular components that users can freely arrange and activate via dedicated controls (Figure~\ref{fig:Teaser}, Region j). At the core of the system is the \textit{Context-KG Visualization} (Figure~\ref{fig:Teaser}, Region c), which generates and displays the visualization based on the user’s question and the diversity setting (Figure~\ref{fig:Teaser}, Region i). The user preferences extracted from the user question are displayed at the bottom of the context visualization (Figure~\ref{fig:Teaser}, Region d). The Context visualization further provides an interactive node query tool (Figure~\ref{fig:Teaser}, Region k), which locates and highlights nodes by name, enabling LLM-referenced entities to be immediately surfaced in the context KG. To prevent clutter, edges are revealed on demand when nodes are selected. Users can interact with the legend to highlight clusters (Figure~\ref{fig:Teaser}, Region l), offering a high-level view of structural patterns. Alongside the context KG visualization, we present three additional components. The \textit{Ontology Visualization} (Figure~\ref{fig:Teaser}, Region g) depicts node types and their ontological distances. The \textit{LLM Insights} component (Figure~\ref{fig:Teaser}, Region f) provides guidance and interpretive support for exploring the context KG visualization. Finally, the \textit{Answer Node Graph Visualization} (Figure~\ref{fig:Teaser}, Region h) presents the direct response to the user’s query (the focus), which typically consists of a small subset of nodes that represent the actual answers which can be validated with the context KG visualization.

\section{Evaluation}
\label{sec:evaluation}

In this section, we evaluate the performance of Context-KG, through quantitative and qualitative analyses.

\subsection{Dataset Selection}

We conducted experiments over three real-world KGs: the TVCG KG, the DBLP KG and the PPOD KG \cite{tu2023interactive}. The TVCG KG captures IEEE Transactions on Visualization and Computer Graphics publications and their metadata. The DBLP KG represents a large-scale bibliographic dataset covering computer science publications. The PPOD KG models connections among persons, projects, organizations, and datasets in regional food systems and conservation in smart foodsheds.

\subsection{Evaluating User Preference Extraction}

We evaluated how accurately the LLM extracts user preferences from natural language queries using GPT-4 Turbo \cite{openai_gpt4_turbo} as the LLM model. We used a human-annotated dataset consisting of 135 questions queried from KGs in different domains such as those shown in Table \ref{tab:user_preferences}. Each query was manually labeled with the four key components: node types of interest, attribute of interest, the value of the attribute of interest, and the connected nodes. These questions ranged from simple to multi-hop queries and served as ground truth. Extraction performance was measured using Accuracy, Precision, Recall, and F1 Score \cite{palacio2019evaluation}. As shown in Table \ref{tab:llm_evaluation}, while some minor mismatches arose from semantically equivalent interpretations of attribute values which were ambiguous even for human annotators.

\begin{table}[ht]
\centering
\caption{Evaluation of user preference extraction using an LLM. Higher values indicate better performance.}
\label{tab:llm_evaluation}
\footnotesize
\setlength{\tabcolsep}{6pt}
\begin{tabular}{|l|c|c|c|c|}
\hline
\textbf{Type} & \textbf{Accuracy} & \textbf{Precision} & \textbf{Recall} & \textbf{F1 Score} \\
\hline
Node types of interest & 96.92\% & 98.44\% & 98.44\% & 98.44\% \\
\hline
Attribute of interest & 88.24\% & 93.75\% & 93.75\% & 93.75\% \\
\hline
\makecell[l]{The value of the node \\[-0.3em] attribute of interest} & 91.04\% & 95.31\% & 95.31\% & 95.31\% \\
\hline
Connected nodes & 97.70\% & 98.27\% & 99.42\% & 98.84\% \\
\hline
\end{tabular}
\end{table}

\subsection{Evaluating Sub Cluster Formations of the Connected Node Types}
In connected node regions, nodes linked to a single interest node naturally form sub-clusters, while those linked to multiple interest nodes appear as pie charts to denote cross-cluster membership. To assess how well the sub-cluster formation is preserved, we applied K-Means and Agglomerative Clustering to the context KG visualization. K-Means captures compact clusters, while Agglomerative Clustering identifies hierarchical or irregular structures. Cluster quality was evaluated using Adjusted Rand Index, Normalized Mutual Information, and Silhouette Score \cite{yin2024rapid, mishra2019fast}. Table \ref{tab:cluster_evaluation} reports these metrics confirming the emergence of sub-clusters for authors and concepts in the TVCG KG for the user query: “Find the papers published in 2018, their authors, and their concepts.”

\begin{table}[t!]
\centering
\caption{Clustering evaluation metrics assessing how well the context KG visualization preserves sub-clusters of connected nodes (authors and concepts) for the query “Find the papers published in 2018, their authors, and their concepts.”}
\label{tab:cluster_evaluation}
\scriptsize % smaller font keeps it in one column
\setlength{\tabcolsep}{2pt} % tighter column spacing
\begin{tabular}{|p{0.9cm}|c|c|c|c|c|c|}
\hline
\multirow{2}{*}{\centering \makecell{Node \\ Type}} 
 & \multicolumn{3}{c|}{K-Means Clustering} 
 & \multicolumn{3}{c|}{Agglomerative Clustering} \\
\cline{2-7}
 & \makecell{Adjusted \\ Rand \\ Index} 
 & \makecell{Normalized \\ Mutual \\ Information} 
 & \makecell{Silhouette \\ Score} 
 & \makecell{Adjusted \\ Rand \\ Index} 
 & \makecell{Normalized \\ Mutual \\ Information} 
 & \makecell{Silhouette \\ Score} \\
\hline
Author  & 1.0 & 1.0 & 0.81 & 1.0 & 1.0 & 0.81 \\
Concept & 1.0 & 1.0 & 0.72 & 1.0 & 1.0 & 0.72 \\
\hline
\end{tabular}
\end{table}

\subsection{Scalability}
We evaluated the scalability of our method using the PPOD KG across increasing graph sizes. Experiments were conducted on a Windows 10 machine equipped with an Intel 12th Gen Core i7-1255U CPU (10 cores, 1.7 GHz) and 16 GB of RAM using Python 3.11.7.For each KG sizes tested, we executed 50 human-authored PPOD KG queries and averaged the results. We measured the clustering time, context layout computation, memory usage, and LLM usage. As summarized in Table \ref{tab:scalability}, all components scale smoothly with graph size: computation times increase gradually, memory usage grows only slightly, and LLM overhead rises moderately. These trends indicate indicate that our framework can handle graphs of increasing size efficiently across all measured dimensions.

\begin{table}[ht]
\centering
\caption{Scalability results across increasing KG sizes. As the number of nodes grows from 100 to 1,000, clustering and layout times increase gradually, memory usage remains stable, and LLM latency and token usage scale moderately.}

\label{tab:scalability}
\resizebox{\columnwidth}{!}{%
\small
\setlength{\tabcolsep}{3pt}

\begin{tabular}{|c|c|c|c|c|c|c|}
\hline
\makecell{\textbf{Number}\\\textbf{of Nodes}} & 
\makecell{\textbf{Clustering}\\\textbf{Time (s)}} & 
\makecell{\textbf{Context KG}\\\textbf{Layout (s)}} & 
\makecell{\textbf{Memory}\\\textbf{Usage (MB)}} & 
\makecell{\textbf{LLM}\\\textbf{Latency (s)}} & 
\makecell{\textbf{LLM Input}\\\textbf{Tokens}} & 
\makecell{\textbf{LLM Output}\\\textbf{Tokens}} \\
\hline
100  & 11.568 & 19.2565 & 734.68 & 16.9299 & 506 & 620  \\
200  & 12.101 & 21.1326 & 738.20 & 17.2457 & 525 & 648  \\
300  & 13.216 & 21.9947 & 740.14 & 17.7204 & 540 & 662  \\
400  & 14.745 & 22.3343 & 741.63 & 18.5693 & 550 & 689  \\
500  & 17.532 & 22.3864 & 742.41 & 18.6965 & 568 & 705  \\
600  & 17.739 & 23.2446 & 743.59 & 19.7309 & 576 & 713  \\
700  & 18.464 & 25.1555 & 744.37 & 20.6727 & 583 & 724  \\
800  & 22.866 & 25.5367 & 744.71 & 21.5821 & 597 & 739  \\
900  & 24.662 & 28.9781 & 746.61 & 22.0552 & 604 & 756  \\
1000 & 26.166 & 29.2399 & 752.62 & 24.5545 & 614 & 771  \\
\hline
\end{tabular}
}%
\end{table}

\subsection{User Study}

\textbf{Participants.} We invited 15 participants: 13 PhD students across the fields of scientific data visualization, information visualization, engineering, and chemistry, one research scientist, and one senior data scientist with extensive KG expertise.

\textbf{Study Setup.} To rigorously assess the effectiveness of Context-KG, we conducted a comparative evaluation against two baseline KG visualization tools: (1) Neo4j and (2) KinGVisher. Because Context-KG uses natural language queries (NLQs) while baselines require SPARQL/Cypher, we generated equivalent baseline queries, ensuring comparison is based solely on visualization quality, not query modality. Participants completed twelve tasks covering both broad exploration and question-driven analysis, summarized in Table \ref{tab:user_study}.

\begin{table}[ht]
\centering
\caption{Overview of KG explorations tasks in the user study} 
\label{tab:user_study}
{\footnotesize
\begin{tabular}{|p{0.95\columnwidth}|}
\hline
\textbf{T1: User Preference Extraction.} Participants specify node types, attributes, and connected nodes in their query and verify whether the visualization reflects these preferences.\\
\hline
\textbf{T2: Attribute-Based Clustering Comprehension.} Participants identify clusters and their shared attribute values to assess how clearly attribute-based grouping is shown.\\
\hline
\textbf{T3: Ontological Grid Awareness.} Participants determine the contents and placement of each node-type region to evaluate how well ontological distances are communicated. \\
\hline
\textbf{T4: Semantic Relationship Discovery} Participants identify spatially close nodes and explain proximity as a reflection of semantic similarity. \\
\hline
\textbf{T5: Cause of Node Placement Interpretation.} Participants select a node and interpret its position and color composition to understand placement rationale. \\
\hline
\textbf{T6: Contextual Understanding.} Participants locate clusters matching their chosen attribute value and compare them with others to evaluate context-driven comparison. \\
\hline
\textbf{T7: Visual Inferencing.} Participants infer outliers, dominant clusters, or bridging roles from various sections of the visualization and justify their inferences based on node colors, positions, and connections. \\
\hline
\textbf{T8: LLM Insights.} Participants evaluate LLM-generated insights and cross-check them with the visualization, assessing their usefulness, accuracy, and trustworthiness in guiding exploration.\\
\hline
\textbf{T9: Node Context.} Participants issue a node-context query and verify how the visualization reflects the added node-level context.\\
\hline
\textbf{T10: Edge Context.} Participants provide an edge-context query and assess whether emphasized relational edges appear as expected.\\
\hline
\textbf{T11: Path Context.} Participants pose a path-context query and examine how the connecting paths are revealed and interpreted.\\
\hline
\textbf{T12: Dynamic Query Refinement and Iterative Exploration.} Participants refine their query and evaluate how well the visualization adapts across iterations. \\
\hline
\end{tabular}
}
\end{table}

\textbf{Procedure.} Each session began with a 15-minute introduction to all three systems, using the TVCG KG. Participants explored the interfaces and asked clarifying questions to ensure they felt confident in using the systems. They then complete all assigned tasks (Table \ref{tab:user_study}) on each system, answering two Likert-scale questions per task \cite{likert1932technique}. In addition to these subjective ratings, we also measured task performance through a three-point task completion score (1 = unsuccessful, 2 = partially successful completion, 3 = fully successful completion) \cite{denisova2024enabling}. Scores were averaged across participants and tasks for system-level comparison. Full task questions are provided in the supplementary material. We also conducted unstructured interviews and collected open-ended feedback on system strengths, limitations, and comparative impressions.

\subsubsection{Results}

\begin{figure}[h]% specify a combination of t, b, p, or h for top, bottom, on its own page, or here
  \centering % avoid the use of \begin{center}...\end{center} and use \centering instead (more compact)
  \includegraphics[width=\columnwidth, alt={User study evaluation results across the nine tasks. Scores are averaged for each task category.}]{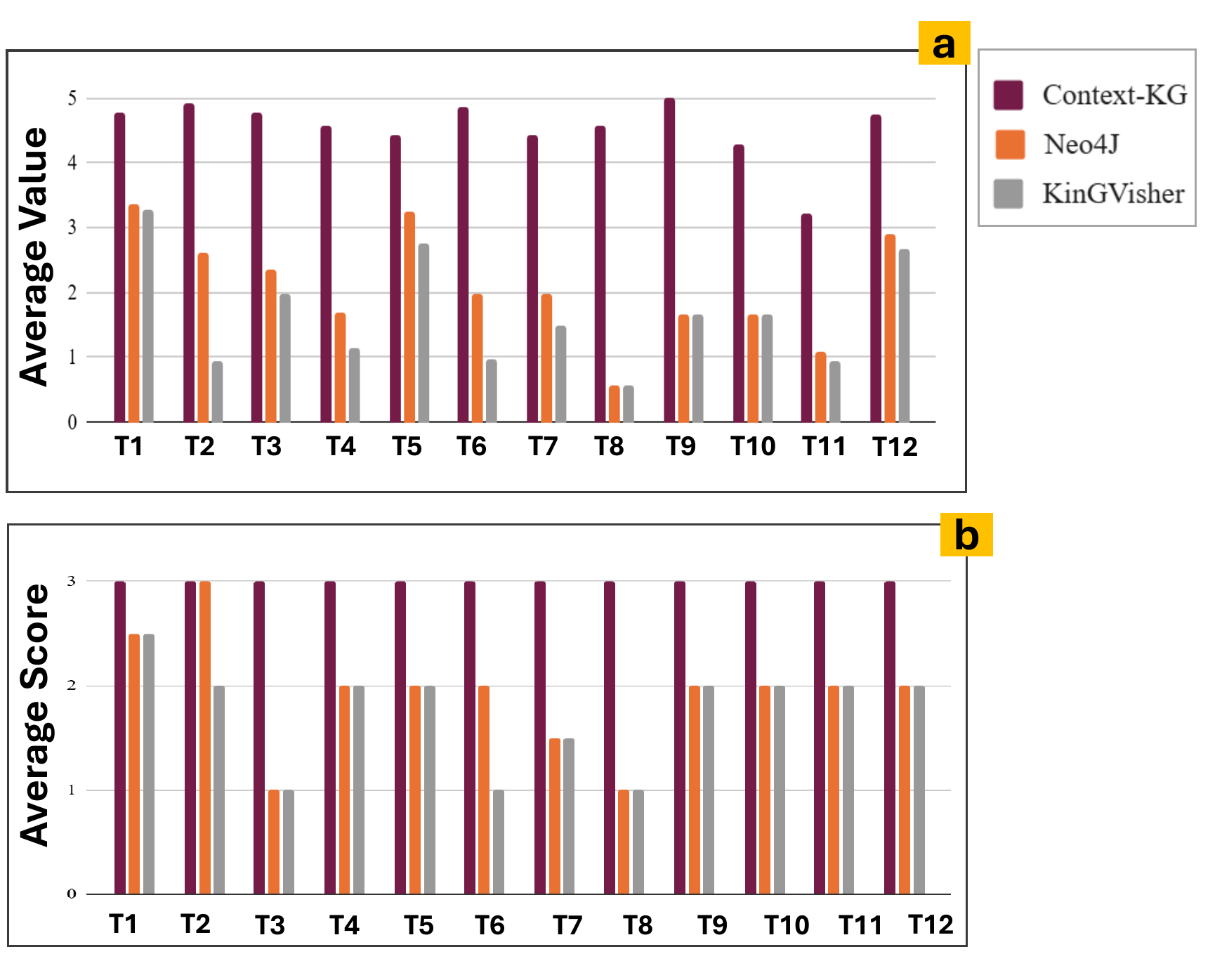}
  \caption{%
  	User study evaluation results across the nine tasks. (a) Average values for each task category scored by users based on the Likert scale and (b) average score based on task completion. %
  }
  \label{fig:user_study_4}
\end{figure}

\textbf{System Ratings.}
Figure~\ref{fig:user_study_4}-a, summarizes average participant ratings across the nine evaluation dimensions for Context-KG, Neo4J, and
KinGVisher. Context-KG achieved the highest ratings across all tasks, with participants emphasizing that its contextual views beyond direct query answers substantially improved KG exploration and query reformulation.

\textbf{Task Completion.}
Across all tasks, Context-KG consistently achieved full task-completion scores (3) as shown by Figure~\ref{fig:user_study_4}-b, whereas Neo4J and KinGVisher frequently received partial or unsuccessful scores—especially on tasks involving cluster interpretation, ontological regions, semantic proximity, and contextual reasoning. These results demonstrate that Context-KG supports reliable multi-step task execution, whereas the baselines frequently hindered task progress due to their lack of contextual structure and semantic guidance.

\textbf{Survey Responses and Feedback.}
We analyzed the average scores across tasks to assess overall user satisfaction. Overall, participants gave positive feedback: for T1, "I liked how my preferences were clearly shown, and how the context visualization was tailored to them." For T2, "The colored polygons made cluster boundaries immediately clear." Users valued the ontological awareness, noting that separating node types into distinct regions made the visualization more intuitive and easier to interpret. Thus for T3, "I found it very easy to distinguish node types thanks to the separate regions, and the layout clearly reflected the ontological distances compared to the cluttered views in Neo4J or KinGVisher." For T4, "It was very clear that nodes placed close together were semantically related, making it easy to understand their connections." T5 feedback included, "Node position and color conveyed cluster membership, and pie charts effectively highlighted key cluster-bridging nodes." For T6, participants said, "Clear clusters and bridging nodes made collaboration patterns and other inferences easy to identify." T7 comments noted, "Context-KG gave more contextual understanding for exploration than other systems." Finally, for T9, users remarked, "The system dynamically updated the visualization to reflect my changing preferences."

The users also suggested potential improvements due to misinterpreted spatial proximity. Such misinterpretations were rare (observed only by 2 participants), and primarily observed in tasks involving dense neighborhoods. Task T8 received a score of 4.22/5, with one user commenting, "The LLM may not capture all the statistical insights I am interested in". Despite these minor limitations, the advantages of Context-KG outweigh these concerns, with users affirming its superior support for efficient and context-rich KG exploration, enhancing user satisfaction compared to the baselines.

\textbf{LLM Insight Evaluation}
To evaluate the reliability and factual accuracy of our LLM-generated insights, we conducted a human evaluation study over 50 diverse user queries from the TVCG KG. Insights were manually checked for hallucinations, incorrect entities or relations, aggregation errors, and faulty inferences and none were observed. The reliability of the results is due to the fact that the LLM is not asked to generate insights blindly; it is provided with carefully encoded structural and relational features of the context KG, including cluster organization, cross-cluster connections, node attributes, and linkages. With this guided input, it produced accurate, evidence-based interpretations for all evaluated queries.

\subsection{Case Studies}
To demonstrate the effectiveness and versatility of Context-KG, we present two representative case studies, showcasing how varying user questions generate distinct, context-aware KG visualizations. We use the TVCG and DBLP KGs for our two case studies.

\begin{figure}[h]% specify a combination of t, b, p, or h for top, bottom, on its own page, or here
  \centering % avoid the use of \begin{center}...\end{center} and use \centering instead (more compact)
  \includegraphics[width=\columnwidth, alt={Use case showing how the modified query “Find papers published in 2016, their authors, and their concepts” supports exploring year-based paper clusters and their relationships to both authors and concepts.}]{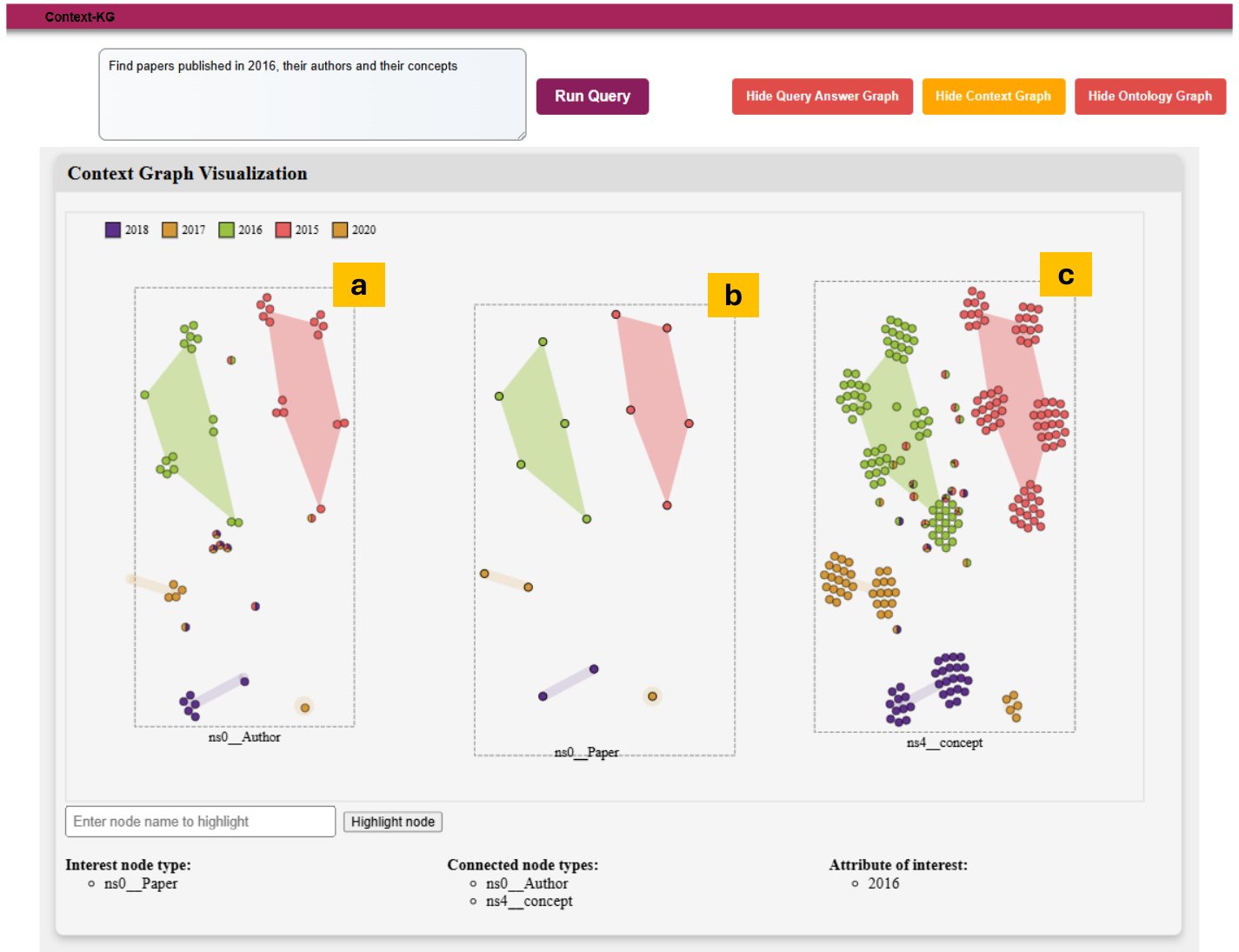}
  \caption{%
  	Use case showing how the modified query “Find papers published in 2016, their authors, and their concepts” supports exploring year-based paper clusters and their relationships to both authors and concepts.%
  }
  \label{fig:use_case_2}
\end{figure}

\textbf{Case Study with the TVCG KG}. In the first case study, a research scientist \textit{S1} explored the TVCG KG with the question: "Find papers published in 2018 and their authors". Initially, answer nodes for papers and authors were displayed (Figure~\ref{fig:Teaser}, Region h). \textit{S1} then transitioned to the Context KG Visualization (Figure~\ref{fig:Teaser}, Region c) to situate the year attribute within its broader context. Using legend-based highlighting, \textit{S1} located the target cluster of 2018 papers and explored surrounding clusters by inspecting node connections. \textit{S1} verified that ontological distances were preserved: Author and Paper nodes were positioned one unit apart, consistent with the ontology (Figure~\ref{fig:Teaser}, Region g). Next, \textit{S1} engaged with the LLM Insights component (Figure~\ref{fig:Teaser}, Region f). This provided interpretive summaries such as highlighting cluster sizes and identifying hubs and bridging nodes, which complemented \textit{S1}'s visual exploration and enhanced understanding of the KG structure relative to the question. Next, \textit{S1} refined the query to “Find the papers published in 2016, their authors, and their concepts” (Figure~\ref{fig:use_case_2}). This reformulation produced a new context-specific KG visualization, illustrating how node arrangement adapts to user intent even within the same KG. The ontology graph remained unchanged for this scenario (Figure~\ref{fig:Teaser}, Region g), since it was derived from the same KG. The updated visualization presented the papers (Figure~\ref{fig:use_case_2}, Region b), authors (Figure~\ref{fig:use_case_2}, Region a), and concepts (Figure~\ref{fig:use_case_2}, Region c) as distinct node-type regions positioned by ontological distance, enabling \textit{S1} to examine primary concepts per paper and detect shared concepts across papers. This dual perspective of the direct answers to the question (the answer nodes and edges) and the surrounding contextual information that enriched the interpretation of the results. In the supplementary material, we provide another use case where \textit{S1} explores papers similar in title (a textual attribute) to a given query paper; this scenario demonstrates how Context-KG supports similarity-based retrieval and cluster navigation for different attribute types.

% add to this case study where we change the question to also show concepts

% \begin{figure*}[htbp]% specify a combination of t, b, p, or h for top, bottom, on its own page, or here
%   \centering % avoid the use of \begin{center}...\end{center} and use \centering instead (more compact)
%   \includegraphics[height=8cm, keepaspectratio, alt={Use case illustrating the context KG visualization for the a user question involving publication titles. (a) depicts the resulting context KG visualization, (b) illustrates the exact query answer, (c) shows the LLM insights and (d) depicts the ontology graph illustrating the ontological distances}]{fig/use_case_1}
%   \caption{%
%   	Use case illustrating the context KG visualization for the a user question involving publication titles. (a) depicts the resulting context KG visualization, (b) illustrates the exact query answer, (c) shows the LLM insights and (d) depicts the ontology graph illustrating the ontological distances%
%   }
%   \label{fig:use_case_1}
% \end{figure*}

\begin{figure}[h]% specify a combination of t, b, p, or h for top, bottom, on its own page, or here
  \centering % avoid the use of \begin{center}...\end{center} and use \centering instead (more compact)
  \includegraphics[width=\columnwidth, alt={The DBLP KG queried with "Show the papers published in 1997 and their authors" resulting in context KG displayed with publication clusters created based on the year of publication.}]{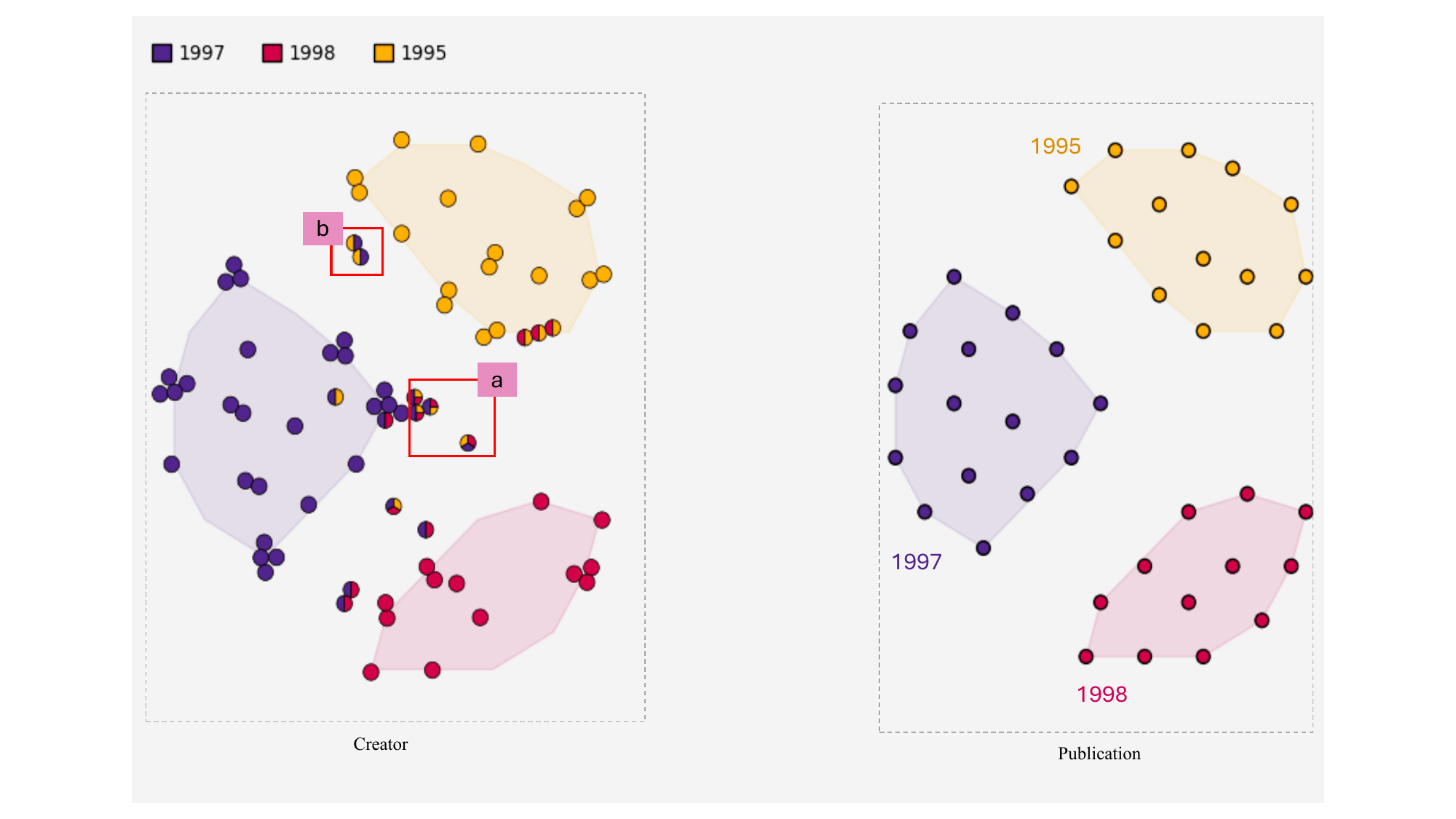}
  \caption{%
  	The DBLP KG queried with "Show the papers published in 1997 and their authors" resulting in context KG displayed with publication clusters created based on the year of publication.%
  }
  \label{fig:use_case_3}
\end{figure}

\textbf{Case Study with the DBLP KG}. The second case study involved a PhD student \textit{S2} with expertise in the visualization domain who sought to explore the DBLP KG to gain insights into publication trends in computer science. Focusing on temporal patterns, \textit{S2} queried: “Show the papers published in 1997 and their authors,” prompting a clustered visualization in which papers were grouped by publication year similar to the TVCG KG. Leveraging this context-aware view, \textit{S2} was able to clearly distinguish clusters of papers and identify their corresponding author groups, thereby uncovering structural patterns in scholarly publishing. The visualization also surfaced key roles, including hub authors linked to many clusters (Figure~\ref{fig:use_case_3}, Region a) and bridging nodes connecting otherwise separate years (Figure~\ref{fig:use_case_3}, Region b). Through iterative querying and reflection, \textit{S2} gained insight into temporal output patterns and cross-community collaboration, moving beyond retrieval toward understanding scholarly dynamics.

\section{Discussion}

In this work, we demonstrated how context-aware, ontology-guided visualization enriched with user preferences can transform traditional KG exploration into a more interpretable and goal-oriented process. By aligning visual structures with user-intended nodes and attributes of interest, our approach reduces the clutter of conventional force-directed layouts and instead surfaces meaningful insights such as attribute-driven clusters, hubs, and bridging nodes. The integration of LLM based insights further complements this visual exploration. Together, these advances show the potential of our system to bridge the gap between raw KG data and actionable insights, paving the way for more intuitive and preference-driven KG analysis.

However, several limitations remain. First, our implementation relies on a single LLM for generating insights, tying evaluation to that model’s capabilities. Future work should examine adaptability across LLMs with varying strengths and prompting requirements. Second, LLM-generated insights are not yet automatically embedded and reflected in the visualization, a capability we aim to integrate. Third, we aim to support a broader range of user questions, not limiting to a primary node type of interest and more complex query structures. Finally, in response to concerns about relational diversity and multi-relational interpretability, we will extend our method to explicitly incorporate edge types, edge attributes, and multi-relational structures into both clustering and layout computations as future work. This includes integrating relation semantics, multiplicity, and attribute-driven edge weighting into the visual encoding, ensuring that relational structure meaningfully contributes to cluster formation and placement. These limitations highlight opportunities to refine the context KG visualization and broaden its applicability across different user contexts.

\section{Conclusion}

This paper introduces Context-KG, an ontology-guided, context-aware visualization framework that enhances user exploration of KGs by integrating semantic structure and user preferences. By moving beyond conventional force-directed layouts, Context-KG situates query results within their ontological and contextual neighborhoods, enabling richer and more meaningful navigation of interconnected data. The combination of ontology-based structuring, user-centric customization, and LLM-driven insights creates an integrated environment where visualization not only reveals relationships but also supports interpretation and reflection. As KGs continue to expand in scale and complexity, approaches like Context-KG will be increasingly vital in aiding users to explore, comprehend, and derive insights from richly connected data. It establishes a powerful visual analytics system for KG exploration in an interpretable and context-sensitive manner.

% \input{Tex_files/10_acknowledgements}

% \section{Appendices}
% % \label{sec:appendices_inst}

% \section*{Supplemental Materials}
% \label{sec:supplemental_materials}

% %% if specified like this the section will be omitted in review mode
% \acknowledgments{%
% 	The authors wish to thank A, B, and C.
%   This work was supported in part by a grant from XYZ (\# 12345-67890).%
% }

\bibliographystyle{abbrv-doi-hyperref}

\bibliography{template}

\appendix % You can use the `hideappendix` class option to skip everything after \appendix
\crefalias{section}{appendix} % this is to make sure that cleverref switches to referring to Appx. X from here on

\end{document}